\documentstyle[aps,prd]{revtex}
\input psfig

\newcommand{\xvec}{{\bf x}}
\newcommand{\phivec}{{\mbox{\boldmath $\phi$}}}
\newcommand{\nvec}{{\bf n}}

\def\xvec{{\bf x}}
\def\kvec{{\bf k}}
\def\rvec{{\bf r}}

\begin{document}

\title{
		     ``Lattice-Free" Simulations of Topological Defect
		     Formation
}
\author{             Robert\ J.\ Scherrer, }
\address{
{\it
		     NASA/Fermilab Astrophysics Center, \\
		     Fermi National Accelerator Laboratory, \\
		     P.O. Box 500, \\
		     Batavia, IL  60510, \\
		     and \\
		     Department of Physics and Department of Astronomy, \\
		     The Ohio State University,               \\
	             Columbus, Ohio 43210                     \\
}}
\author{             Alexander Vilenkin        }
\address{
{\it
                     Institute of Cosmology,
                     Department of Physics and Astronomy,     \\
                     Tufts University, Medford, MA 02155        \\
}}

\maketitle
%

\renewcommand{\baselinestretch}{1.3}
\begin{abstract}

We examine simulations of the formation of domain walls, cosmic
strings, and monopoles on a cubic lattice, in which the topological defects
are assumed to lie at the zeros of a piecewise constant 1, 2, or 3 component
Gaussian random field, respectively.  We derive analytic
expressions for the corresponding topological defect densities
in the continuum limit and show
that they fail to agree with simulation results, even
when the fields are smoothed on small scales to eliminate
lattice effects.  We demonstrate that this discrepancy,
which is related to a classic geometric fallacy, is due to
the anisotropy of the cubic lattice, which cannot be eliminated
by smoothing.  This problem can be resolved by linearly interpolating
the field values on the lattice, which gives results in good agreement
with the continuum predictions.  We use this procedure to
obtain a lattice-free estimate (for Gaussian smoothing)
of the fraction of the total length of string
in the form of infinite strings: $f_\infty = 0.716 \pm 0.015$.
\end{abstract}
\pacs{PACS numbers: }
%


\section{Introduction}

Topological defects arise when the
manifold of the
degenerate vacuum states
of a field is not simply connected.
(See Ref. \cite{review} for a review).  In the
simplest set of models, domain walls arise from
a one-component field with two degenerate vacuum
states, cosmic strings correspond to a two-component
field in which the degerate vacua form a circle
in the field space, and monopoles arise from a three-component
field for which the degenerate vacua lie on the surface of a sphere.

The formation of topological defects can
be simulated by laying down the appropriate field on
a cubic lattice, and identifying the corresponding
topological defects with zeros of the field.
Although the most effort has gone toward simulations
of cosmic string formation \cite{VV} - \cite{Borrill},
similar investigations have also been undertaken
for the study of domain walls \cite{Press} - \cite{Larsson} and
monopoles \cite{Copeland} - \cite{Bennett}.

In this paper, we point out a potential problem
for some such lattice-based simulations.
Even when the fields which give rise to the defects
are smoothed on a scale larger than the lattice
spacing (as in Ref. \cite{paper1}), residual
lattice effects remain.  These are due
to the fact that a continuous surface cannot be distorted
to lie on the edges of a lattice without also distorting
the area of the surface.

In the next section, we discuss our topological
defect simulations, and we provide analytic expressions
for the density of walls, strings, and monopoles.  These
analytic predictions are compared with the results of our numerical
simulations, and they are found to disagree.
In Sec. 3,
a second set of analytic predictions, which does agree with
the simulations, is presented, and we explain the origin
of the discrepancy, which is closely related to a classic
geometric fallacy.  A second set of simulations using
linear interpolation of the fields is presented, and these
are shown to agree with analytic continuum predictions.
The implications are discussed
in Sec. 4.  Our interpolation scheme produces a lattice-free
estimate for the fraction of string length in infinite strings
(with Gaussian smoothing):  $f_\infty = 0.716 \pm 0.015$.

\section{The Problem}

In an earlier paper \cite{paper1} we simulated
the formation of cosmic strings from correlated
fields on a cubic lattice, smoothing the field on small scales
to eliminate lattice effects.  Here we extend
these simulations to domain walls and monopoles.

Consider first the case of domain walls.
We can assign a real scalar field
$\phi$ to each of the
cells of a cubic lattice, and the location of the
domain wall is then identified with faces of the
lattice across which the value of the field
crosses zero.  To simulate cosmic strings rather than
domain walls, we use a complex scalar
field, or equivalently, a two-component
real field $\phivec$.  Then the zeros of $\phivec$ will
lie on the edges of the lattice, which we then
identify as the location of the strings.
Finally, to simulate monopoles, we take $\phivec$ to
be a three-component real field; the zeros of $\phivec$
then form a set of disconnected points lying on the
vertices of the lattice; this gives the location of the
corresponding monopoles.

Following the procedure in Ref. \cite{paper1}, we take
$\phivec$ to be a Gaussian random field.
Then we can exploit the fact that the components
of an $N-$component Gaussian field, $\phi_1, \phi_2,... \phi_N$,
are themselves
independent real Gaussian fields.  Hence, our
model for cosmic strings is equivalent to using
two independent real Gaussian fields (which can be taken
to be $\phi_1$ and $\phi_2$).  The zeros
of these two fields form two independent sets
of surfaces, and the strings lie at the intersection
of these surfaces.  Similarly, our field configuration for monopoles
can be treated as three independent real Gaussian fields,
$\phi_1$, $\phi_2$, and $\phi_3$.
The zeros of these three fields form three sets of
surfaces, and their intersection is a set of disconnected
points, giving the location of the monopoles.

A Gaussian field is completely determined by its power spectrum
$P(k)$,
defined by
\begin{equation}
P(k) = \int d^3 r~ e^{i \kvec\cdot \rvec} \langle \phivec(\xvec)
\cdot \phivec(\xvec+\rvec)\rangle.
\end{equation}
Following the procedure in Ref. \cite{paper1}, we assume
a power-law power spectrum of the form
\begin{equation}
\label{power}
P(k) \propto k^n.
\end{equation}
Note that this is equivalent to taking each of the components
$\phi_i$ to have a power spectrum given by equation (\ref{power}).
We add a backround non-zero mean field to the box
to simulate the effect of long-wavelength modes
(see Ref. \cite{paper1} for the details).  We also
smooth the field $\phivec$ on small scales with a Gaussian
window function
\begin{equation}
\label{gauss}
W(r) = \exp(-r^2 / 2 r_0^2),
\end{equation}
where the smoothed field
$\phivec_s(\xvec)$ is the convolution of $\phivec(\xvec)$ with $W(\rvec)$:
\begin{equation}
\label{phismooth}
\phivec_s(\xvec) = \int d^3 r ~\phivec(\xvec+\rvec) W(\rvec).
\end{equation}
This Gaussian smoothing changes the power spectrum from
$P(k)$ to $P_s(k)$, given by
\begin{equation}
\label{powersmooth}
P_s(k) = P(k) e^{-k^2 r_0^2}
\end{equation}
(Again, equation (\ref{phismooth}) is equivalent to smoothing
each of the components $\phi_i$ with $W(\rvec)$, and
equation (\ref{powersmooth}) can similarly be applied
separately to the power spectrum for each component.)
The effect of smoothing is to reduce the magnitude of the small-scale
fluctuations by averaging them out over the window function.
One might hope that if $r_0$ is taken to be larger than the
lattice spacing, lattice effects can be reduced or eliminated
entirely.  We will see that this is not the case.

Given this model, it is possible to calculate the 
topological defect density analytically.  Our starting
point is a set of results due to Ryden \cite{Ryden}
concerning the properties of level-crossing surfaces
in Gaussian random fields.  Consider a real Gaussian field
with arbitrary power spectrum $P(k)$ and rms
fluctuation $\sigma$.  The regions of space
for which the field has the value $\nu \sigma$
form a set of two-dimensional surfaces.  Ryden
\cite{Ryden} defined the quantity $N_3(\nu)$
to be the mean area per unit volume of these surfaces.
The intersection of this set of surfaces with an arbitrary
plane produces a set of curves, with mean length $N_2(\nu)$
per unit area of the plane.  Finally, we can consider
the intersection of this set of surfaces with an arbitrary
straight line; the quantity $N_1(\nu)$ is the number
of intersections of this line (per unit length of the line)
with the set of surfaces.  Ryden showed
that \cite{Ryden}
\begin{equation}
N_3(\nu) = {4 \over \pi} N_2(\nu) = 2 N_1(\nu),
\end{equation}
where
\begin{equation}
N_1(\nu) = {1 \over \pi \sqrt{3}} \langle k^2 \rangle^{1/2} e^{-\nu^2 /2},
\end{equation}
and the value of $\langle k^2 \rangle$ depends on the power spectrum
as
\begin{equation}
\label{k2}
\langle k^2 \rangle = {\int P(k) k^4 dk \over  \int P(k) k^2 dk}
\end{equation}

These results can be used to derive analytic expressions for
the topological defect density in our model.
Consider first the area per unit volume, $(A/V)$, for
domain walls.  In terms of the Ryden notation, it is obvious
that
\begin{equation}
\label{(A/V)}
(A/V) = N_3(0) = {2 \over \pi \sqrt{3}} \langle k^2 \rangle^{1/2}
\end{equation}
(Although we confine our attention to unbiased field configurations,
i.e., configurations for which positive and negative values
of $\phi_i$ are equally likely,
so that $\nu = 0$, our results can easily be generalized
to biased configurations by taking a nonzero value for $\nu$).
Now consider the case of cosmic strings.  The strings can
be considered to lie at the intersection of two sets of
$\nu=0$ surfaces, corresponding to $\phi_1$ = 0 and
$\phi_2$ = 0, respectively.  Suppose that we are sitting
on a surface corresponding to $\phi_1$ = 0, with
area per unit volume $N_3(0)$,
and we wish
to calculate the length per unit area of the intersection
of this surface with the $\phi_2$ = 0 surface.
Although the $\phi_1$ = 0 surface is highly irregular,
if we take a small enough patch, it will be locally
flat (assuming the field is smoothed on small scales),
so that the length per area of this intersection is just $N_2(0)$.
Hence, the total length per unit area, $(L/V)$, of cosmic
string is just
\begin{equation}
\label{(L/V)}
(L/V) = N_2(0) N_3(0) = {1 \over 3 \pi} \langle k^2 \rangle
\end{equation}
Note that this result agrees exactly with the analytic expression
for $(L/V)$ derived by Vishniac et al. \cite{VOS} using
different methods.
Finally, we consider the case of monopoles, which are taken
to lie at the intersection of the
surfaces corresponding to the zeros of three independent Gaussian fields.
The intersection of two of these sets of surfaces ($\phi_1 = 0$
and $\phi_2$ = 0) defines a set of curves
with length per unit volume given by equation (\ref{(L/V)}).
While these curves are quite irregular, they are locally straight
on small enough scales (assuming the fields are smoothed
on small scales), so that their intersection with the third
set of surfaces ($\phi_3 = 0$) gives $N_1(0)$ as the number of intersections
per unit length of the curves.  Then the monopole density
per unit volume, $(N/V)$, is
\begin{equation}
\label{(N/V)}
(N/V) = N_1(0) N_2(0) N_3(0) = {1 \over 3^{3/2} \pi^2} \langle k^2 \rangle^{3/2}
\end{equation}

Although our results are applicable to arbitrary power spectra,
for definiteness we now restrict our attention to the case
$n=0$, which corresponds to uncorrelated fields.
This is the most physically relevant case, because a causal phase
transition will lead to fields which are uncorrelated on
scales larger than the horizon, while the smoothing length
can be identified with the scale over which the field tends
to be correlated.
For $n=0$ with Gaussian smoothing, equations (\ref{power}),
(\ref{powersmooth}), and (\ref{k2}) yield
\begin{equation}
\langle k^2 \rangle = {3 \over 2 r_0^2}
\end{equation}
Substituting this result into equations (\ref{(A/V)}), (\ref{(L/V)})
and (\ref{(N/V)}), we obtain
\begin{eqnarray}
\label{numtheory}
(A/V) &=& {\sqrt{2} \over \pi r_0}~~~ ~~~~~~~~~({\rm walls}),\nonumber\\
(L/V) &=& {1 \over 2\pi r_0^2} ~~~~~~~~~~({\rm strings}),\nonumber\\
(N/V) &=& {1 \over 2^{3/2} \pi^2 r_0^3}  ~~~~~({\rm monopoles}).
\end{eqnarray}

We now compare these results to numerical simulations.
Using the procedure outlined above, we have simulated
the formation of domain walls, cosmic strings, and monopoles
on a $128^3$ lattice, using 1, 2, and 3 independent Gaussian
random fields, respectively.  For the case of cosmic strings,
we use two different cubic lattices, staggered with respect
to each other by 1/2 of a lattice spacing in the $x$, $y$,
and $z$ directions (i.e., the vertices of one lattice lie
at the centers of the cells of the other lattice).  We then
set down independent Gaussian fields on each lattice,
and the strings are taken to lie at the intersection of the zeros of the two
fields.  For monopoles, we take three different cubic
lattices, staggered with respect to each other by 1/3 of a lattice
spacing in the $x$, $y$, and $z$ directions, and the monopoles are taken to
lie at the intersection of the three surfaces defined
by $\phi_1 = 0$, $\phi_2 = 0$, and $\phi_3 = 0$.

In each case, we produce
4 different realizations to obtain a mean defect number density
and standard deviation.
In Fig. 1, we show the domain wall area per unit volume, (A/V),
in our simulations as a function of smoothing length $r_0$.
In Fig. 2 we give the string length per unit volume, (L/V), as
a function of $r_0$, and
in Fig. 3 we give the number of monopoles per unit volume,
$(N/V)$, as a function of $r_0$.
(Here we make no distinction between monopoles and antimonopoles,
although we will do so later).
Also in these figures, we show as solid lines the theoretical values
for $(A/V)$, $(L/V)$, and $(N/V)$ from equations ($\ref{numtheory}$).
The disagreement is obvious.  Clearly the lattice-based simulations
fail to produce the true continuum number densities for topological
defects.

\begin{figure}
\hspace*{1.3in}
\psfig{file=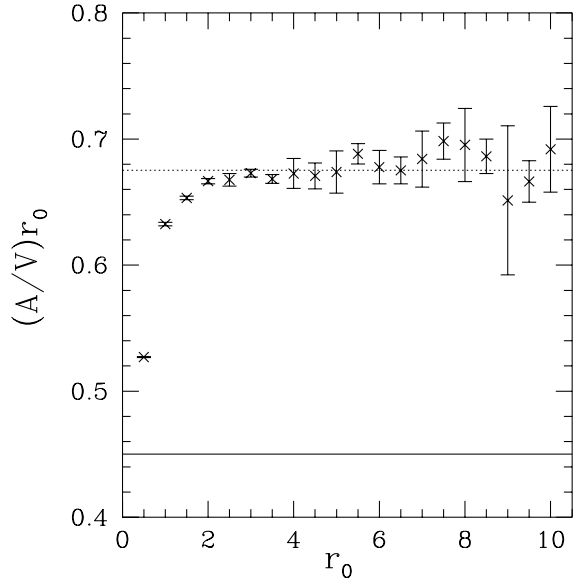,height=8.cm,width= 8.cm}
\vspace*{0.4in}
\caption{The total domain wall area per unit
volume, $(A/V)$, multiplied by the smoothing length $r_0$,
as a function of $r_0$, for Gaussian smoothing,
where $r_0$ is measured in units of the lattice spacing.
The solid line gives the analytic prediction for
$(A/V)r_0$ for this model in the continuum limit.
The dotted line is the lattice-based analytic prediction.}
\end{figure}

\begin{figure}
\hspace*{1.3in}
\psfig{file=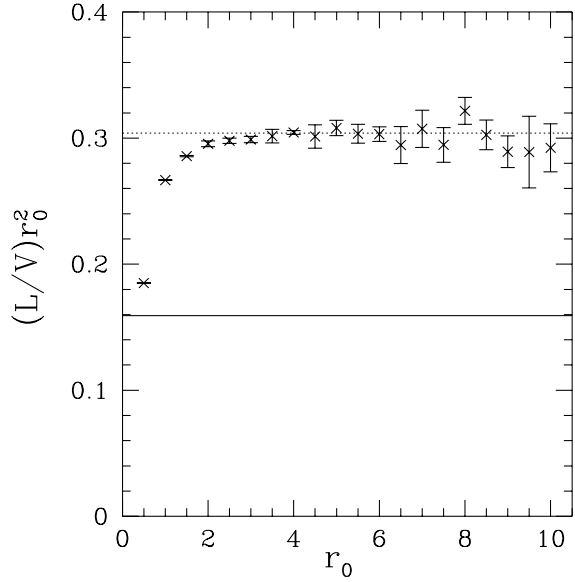,height=8.cm,width= 8.cm}
\vspace*{0.4in}
\caption{The total string length per unit
volume, $(L/V)$, multiplied by the square of the smoothing length $r_0$,
as a function of $r_0$, for Gaussian smoothing, where $r_0$ is
measured in units of the lattice spacing.
The solid line gives the analytic prediction for
$(L/V)r_0^2$ for this model in the continuum limit.
The dotted line is the lattice-based analytic prediction.}
\end{figure}

\begin{figure}
\hspace*{1.3in}
\psfig{file=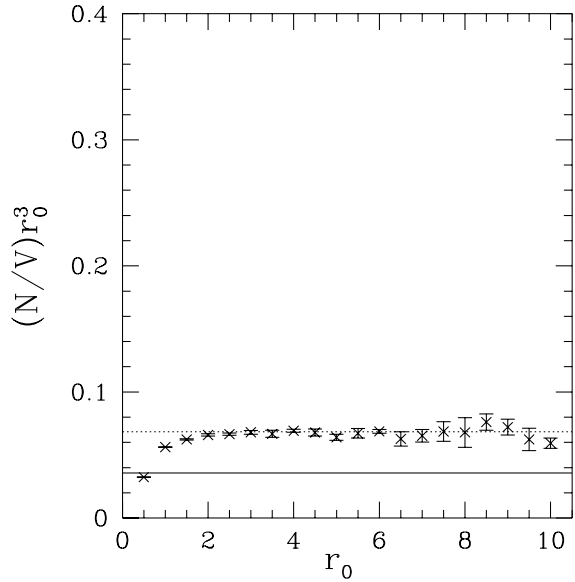,height=8.cm,width=8.cm}
\vspace*{0.4in}
\caption{The total number of monopoles per unit volume,
$(N/V)$, multiplied by the cube of the smoothing length $r_0$,
as a function of $r_0$, for Gaussian smoothing,
where $r_0$ is measured in units of the lattice spacing.
The solid line gives the analytic prediction for
$(N/V)r_0^3$ for this model in the continuum limit.
The dotted line is the lattice-based analytic prediction.}
\end{figure}

\section{The Solution}

Although our analytic expressions for the topological
defect densities given in the previous section fail
to agree with the results of our numerical simulations,
we can use a different set of arguments
to derive ``lattice-based" analytic expressions which do agree
with the lattice simulations.
Consider first the case of domain walls, and consider
a single face between two cells on the lattice.  This face will be occupied
by a domain wall if the cell on one side of it has $\phi >0$ and the other
cell has $\phi < 0$, or vice versa.  This occupation probability,
$p_{occ}$, can be derived
from Sheppard's theorem \cite{Sheppard}
when $\phi$ is a Gaussian
field.   We obtain
\begin{equation}
p_{occ} = {1\over \pi} \cos^{-1}\biggl[{\xi(r) \over \xi(0)}\biggr],
\end{equation}
where $r$ is the spacing between the lattice cells, and $\xi$
is the two-point correlation function, which is the Fourier
transform of the power spectrum.
For an $n=0$ power spectrum
smoothed with our Gaussian window function, we have
$\xi(r)/\xi(0) = \exp(-r^2/4r_0^2)$, and
\begin{equation}
p_{occ} = {1\over\pi} \cos^{-1}\biggl[e^{-r^2 / 4 r_0^2}\biggr].
\end{equation}
For $r_0 \gg r$ (i.e., when the smoothing length is
large compared to the lattice spacing) this reduces
to
\begin{equation}
\label{pocc}
p_{occ} = {1 \over \pi \sqrt{2}r_0}
\end{equation}

This result can be used to derive not only the domain
wall density, but the string and monopole densities as well.
For the case of domain walls, each cell is bounded by
six faces, each of which is shared between two cells,
so that the total area per unit volume is
$(A/V) = 3 p_{occ}$.  In our string simulation, a
string forms at the intersection of two different lattice
faces, where the two lattices are staggered by half a lattice
spacing.  For a single cell on one lattice,
there are 24 such possible intersections, each
with length $r/2$ (where $r$ is the lattice spacing) and
each shared by two cells.  Hence, the total length
per unit volume is $(L/V) = 6 p_{occ}^2$.  Finally,
for the monopoles, a monopole forms at the intersection
of three lattice faces on three different
staggered lattices.  On a single cell of one lattice,
there are 12 possible intersection sites, each of which
is shared between two cells.  Thus, $(N/V) = 6 p_{occ}^3$.

Combining these results with the value for $p_{occ}$
given in equation (\ref{pocc}), we obtain the following
values for the topological defect densities:
\begin{eqnarray}
\label{numlat}
(A/V)_{lat} &=& {3 \over \pi \sqrt{2}r_0} ~~~~~~~({\rm walls}),\nonumber\\
(L/V)_{lat} &=& {3 \over \pi^2 r_0^2} ~~~~~~~~~({\rm strings}),\nonumber\\
(N/V)_{lat} &=& {3 \over \pi^3 \sqrt{2}r_0^3} ~~~~~({\rm monopoles}),
\end{eqnarray}
where the $lat$ subscript denotes the fact that this analytic
calculation is lattice-based, rather than derived from a continuum
calculation.

These values for the defect densities are shown in Figs. 1-3 as
dotted lines.  As $r_0$ becomes sufficiently large compared
to the lattice spacing, the numerical results converge
to a defect density in good agreement with equations (\ref{numlat}),
although the fluctuations
between realizations also increase with $r_0$.
Clearly, our results in equations (\ref{numlat}) give
the correct defect densities to compare with our lattice-based
simulations.

However, this leaves an embarrassing question:  why
do our simulations fail to agree with the continuum calculations
given in equations (\ref{numtheory}) and in reference \cite{VOS}?
Even without reference to the numerical simulations, we
have two predictions for the defect densities,
i.e. equations (\ref{numtheory}) and (\ref{numlat}), which
disagree with each other.  What is the origin
and significance of this discrepancy?

In fact, the problem here is related to a classic
geometric fallacy \cite{fallacy}.  Consider
an isosceles right triangle with sides of unit length.
The hypotenuse can be represented as a series of
steps (Fig. 4).  The length of the hypotenuse,
represented in this way, is 2.  As the number
of steps $N$ is increased, this length remains 2,
even in the limit where $N \rightarrow \infty$.
The fundamental problem is that the hypotenuse is not
the $N \rightarrow \infty$ limit of the zigzag curve.
Similarly, it is impossible to represent accurately
a smooth surface or curve on a cubic lattice.
While smoothing increases the length scale of curvature
for the smooth surface (or, as in Fig. 4, decreases
the effective step size), the lattice representation
of the surface never reaches the continuum limit.

\begin{figure}
\hspace*{1.3in}
\psfig{file=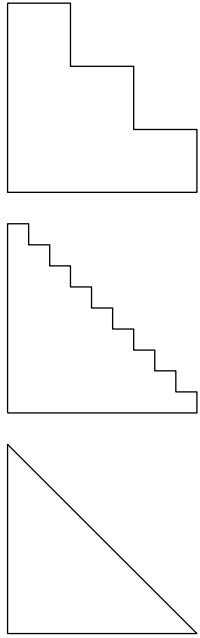,height=10.cm,width= 10.cm}
\vspace*{0.4in}
\caption{An illustration of a classic geometric
fallacy:  as the number of steps goes to infinity, the
length of the hypotenuse remains at 2, rather than approaching
the correct value of $2^{1/2}$.}
\end{figure}

Armed with this information, we can now calculate
the expected ratio of the continuum value for $(A/V)$
given in equation (\ref{numtheory}) and the lattice
value in equation (\ref{numlat}).  Suppose that
we represent a locally flat surface with area $A$ and normal vector
$\nvec$ on a cubic lattice.  The area $A$ then forms
a triangular face of a pyramid with sides of length
$x$, $y$, and $z$.  The volume of the pyramid
is $V = {1 \over 3} An = {1 \over 6} xyz$, so
that $A = {1 \over 2} xyz/n$.  The lattice value
for this surface area is just $A_{lat}={1\over 2}(xy + xz + yz)$.
Then $(A/V)_{lat} / (A/V)$ is just the mean value
of the ratio of $A_{lat}$ to $A$, which is
\begin{eqnarray}
{(A/V)_{lat} \over (A/V)} &=& \biggl\langle{A_{lat} \over A}\biggr\rangle,
\nonumber\\
&=& \biggl\langle{{1 \over 2}(xy+yz+xz) \over {1\over2}
xyz/n}\biggr\rangle,\nonumber\\
&=& \langle \cos \alpha + \cos \beta + \cos \gamma \rangle,
\end{eqnarray}
where $\alpha$, $\beta$, and $\gamma$ are the angles between
$\nvec$ and the $x$, $y$, and $z$ axes.  Since $\nvec$ is
distributed isotropically, we have $\langle \cos\alpha \rangle =
\langle \cos\beta \rangle = \langle \cos\gamma \rangle = 1/2$,
and $(A/V)_{lat} / (A/V) = 3/2$, which is exactly
the value obtained by comparing equation (\ref{numtheory})
to equation (\ref{numlat}).

In fact, the solution to this problem
for the case of domain walls was noted by
Press et al. \cite{Press} and has been used in all domain wall simulations
\cite{Press} - \cite{Larsson}.  In the Press, et al., procedure,
the area of a domain wall is simply multiplied by the factor
$1/( |\cos \alpha| + |\cos \beta| + |\cos \gamma|)$, resulting
in the correct continuum value for the domain wall area.

A similar weighting procedure cannot be applied in
a straightforward way to simulations of cosmic strings and
monopoles.
For strings, a similar calculation yields
$(L/V)_{lat}/(L/V) = \langle (x + y + z)/\sqrt{x^2+y^2+z^2}\rangle = {3\over 2}$,
while the actual ratio from equations (\ref{numtheory})
and (\ref{numlat}) is $(L/V)_{lat}/(L/V) = 6/\pi = 1.91$, which is
slightly larger.  For monopoles, our geometrical argument fails
completely.  There is no obvious reason that $(N/V)$ should be any
different in a continuum or lattice calculation, since the monopoles
are geometrical points.  Nonetheless, equations (\ref{numtheory})
and (\ref{numlat}) yield $(N/V)_{lat}/(N/V) = 6/\pi$, as in the case
of strings.

This secondary lattice dependence is a result of the way
in which we have simulated the string and monopole formation.
Both the strings and monopoles are treated as arising from
intersections of the surfaces defined by the zeros of 2 or 3
independent Gaussian fields.  Even when smoothed, these surfaces
retain a residual jaggedness, as we have already emphasized.
The intersection of two such jagged surfaces produces spurious
additional intersections which would not be present in the continuum
limit.  As an example, consider the hypotenuse in Fig. 4.
When represented as a series of steps, two such lines can
have a large number of intersections, even when the two
lines are parallel, while in the continuum limit, there would be at
most one intersection, and none in the case where the lines
were parallel.
This secondary effect does not enter into
the domain wall calculation, since the domain walls arise
from the zeros of a single field.

This problem can be resolved by linearly interpolating
the values of each field on the lattice, effectively representing
the zeros of the field as polyhedral, rather than cubic,
surfaces.  We have tested such a procedure for the case
of monopoles.  We derive an equation for the value
of $\phi$ for each of the three fields
in each cubic cell by linearly interpolating on the values
of each field at the four points $(0,0,0)$, $(1,0,0)$, $(0,1,0)$,
and $(0,0,1)$.  We then place a monopole in the cell if the
$\phi=0$ planes of the three fields intersect inside the cell.
The results are shown in Fig. 5.  They agree extremely well
with the continuum predictions, despite the fact that our linear
interpolation is still
an approximation to the true continuum field.

\begin{figure}
\hspace*{1.3in}
\psfig{file=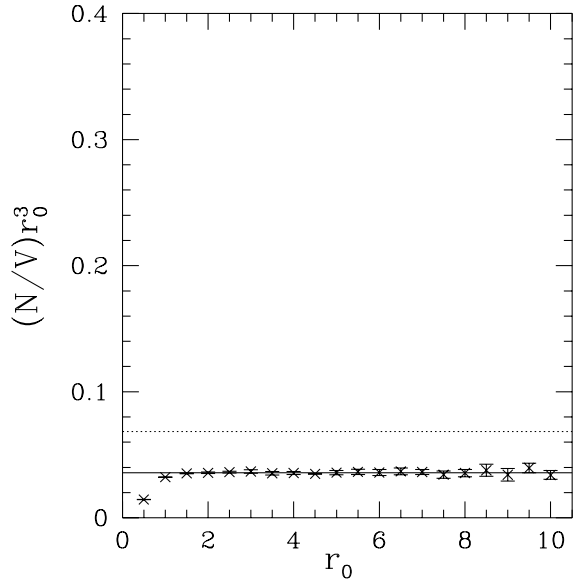,height=8.cm,width= 8.cm}
\vspace*{0.4in}
\caption{As in Fig. 3, but with linear interpolation on the
three fields used to determine the monopole positions.}
\end{figure}

This procedure can also be used for cosmic strings.
In this case, we take the values of our two fields to lie on the vertices
(rather than inside the cells) of the cubic lattice.
For each face of the cubic lattice, and for each of the two fields,
we use
bilinear interpolation \cite{NR} on the four
values of the field bounding that face
to calculate the location of the zeros of the field
on that face.  This gives
the intersection (if any) of the two surfaces $\phi_1 = 0$ and $\phi_2 = 0$
with the given lattice face.  If both surfaces do
intersect a given lattice face, then we obtain two
curves (one for each field) which give the intersection of the $\phi_1 = 0$
and $\phi_2 = 0$
surfaces of
that field with the lattice face.
The intersection of these two curves
gives the point at which a string passes through that face of the cubic
lattice.
We then connect these intersection points with straight segments
(now no longer parallel to the lattice edges)
to obtain the string network.
(The general expression for the location of a single $\phi=0$
curve on a lattice face, obtained from bilinear interpolation,
is $axy + bx + cy + d = 0$, where $a - d$ are constants
and the face is taken to lie in the $x - y$ plane.
Hence, this procedure can produce two intersections of the
$\phi_1 = 0$ and $\phi_2 = 0$ curves on a single lattice face,
corresponding to two
strings passing through a single face of the lattice.
In this case we treat the string as a single small
loop with length equal to twice the distance between
the intersection points.  Such tiny loops become relatively less
important as the smoothing length increases).
The resulting
values for $L/V$ as a function of smoothing length are displayed
in Fig. 6.  Again, agreement with our continuum predictions
is excellent.

\begin{figure}
\hspace*{1.3in}
\psfig{file=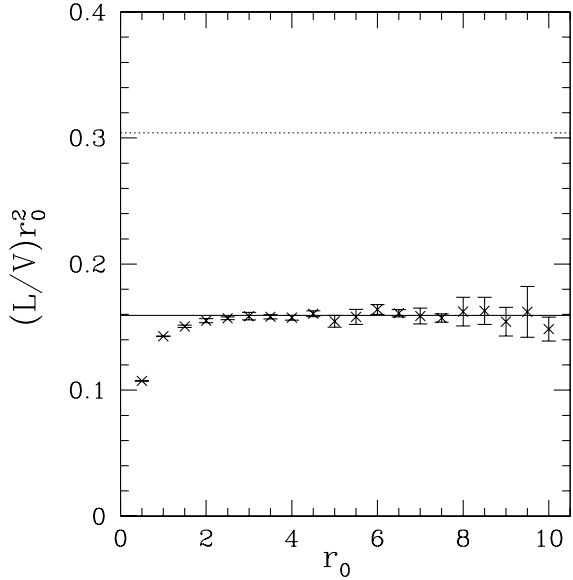,height=8.cm,width= 8.cm}
\vspace*{0.4in}
\caption{As in Fig. 2, but with linear interpolation on the two
fields used to determine the cosmic string positions.}
\end{figure}

We can use this linearly-interpolated string network to calculate
a ``lattice-free" estimate of $f_\infty$, the fraction of string length
in the form of infinite strings, which we define,
as in Ref. \cite{paper1}, to be strings which cross
the entire simulation volume in either the $x$, $y$, or
$z$ direction.  We have calculated $f_\infty$ as
a function of smoothing length for our linearly-interpolated
string simulations.  The results are given in Fig. 7.

\begin{figure}
\hspace*{1.3in}
\psfig{file=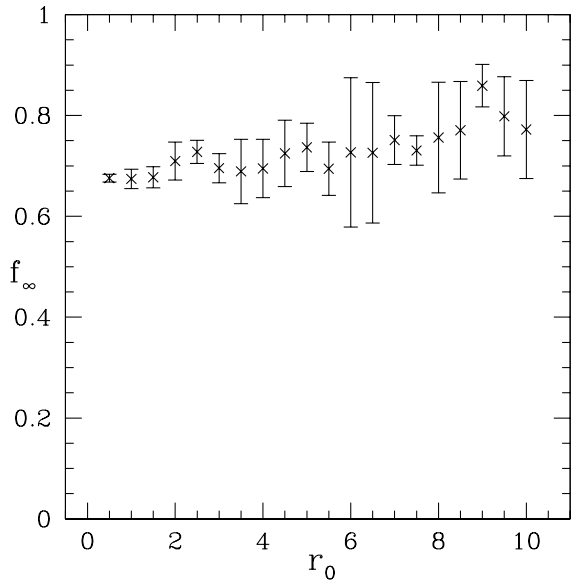,height=8.cm,width= 8.cm}
\vspace*{0.4in}
\caption{The fraction of total string length in the form of
infinite strings, $f_\infty$, as a function of smoothing length,
$r_0$, for Gaussian smoothing, where linear interpolation on the two fields
is used to determine the cosmic string positions.}
\end{figure}

To obtain an estimate of the lattice-free value of
$f_\infty$, we want to choose values
of $r_0$ which are large enough compared to the lattice spacing
to eliminate lattice effects,
but small enough compared to the total box size to avoid
large fluctuations from one run to the next.  The results
shown in Fig. 6 suggest that the range $3 \le r_0 \le 7.5$ should
be suitable.  When we perform a weighted average of the values
of $f_\infty$ in this range, we obtain:  $f_\infty = 0.716 \pm 0.015$.
This result represents a true lattice-free estimate of
$f_\infty$, but we do expect it to depend on
the actual window function used for smoothing.  Our
lattice-free results are
strikingly similar to earlier results
obtained with Gaussian smoothing in a simulation
where strings were constrained to lie on a cubic lattice:
$f_\infty = 0.71 \pm 0.01$ \cite{paper1}.  It would
appear that $f_\infty$, unlike $(L/V)$, is
not much affected by the use of a cubic lattice.

For the case of monopoles, one interpretation of our results
is that our simple-minded lattice simulation leads to the formation of spurious
closely-spaced monopole-antimonopole pairs.  To test this
hypothesis, we must distinguish between monopoles
and antimonopoles in the simulation.  Consider first
the case where the $\phi_1=0$ surface lies in the $y-z$
plane, the $\phi_2 = 0$ surface lies in the $x-z$ plane,
and the $\phi_3 = 0$ surface lies in the $x-y$ plane,
so that the only non-zero derivatives of
the field are $\phi_{1,x}$, $\phi_{2,y}$ and $\phi_{3,z}$.
In this case, we define zeros of the three fields to be monopoles
if $\phi_{1,x} \phi_{2,y} \phi_{3,z} > 0$, and antimonopoles
if this product is negative.  For other field configurations,
the sign of the product of the derivatives which
corresponds to a monopole changes each
time we alter the direction of one pair of zeros.
For instance, if the $\phi_1=0$ surface lies in the $x-z$
plane and the $\phi_2=0$ surface likes in the $y-z$ plane,
but the $\phi_3 = 0$ surface remains in the $x-z$ plane,
then $\phi_{1,y} \phi_{2,x} \phi_{3,z}$ is now negative
for monopoles and positive for antimonopoles.  Similarly,
if we now permute the directions of the zeros of
the $\phi_2$ and $\phi_3$ fields, then
$\phi_{1,y} \phi_{2,z} \phi_{3,x}$ is positive for monopoles
and negative for antimonopoles.

Using this definition of monopoles and antimonopoles, we have checked
the consequences of allowing closely-separated
monopole-antimonopole
pairs in our simulation to annihilate.  We use
the same monopole simulation as in Fig. 3, but now for
each monopole in the simulation, we check for the existence
of any antimonopoles within an ``annihilation distance" $r_A$.
If an antimonopole is found within this distance, both
the monopole and antimonopole are removed from the simulation.
Thus, in our final configuration, all remaining
monopoles are separated
from antimonopoles by a minimum distance of $r_A$.
We have fixed the smoothing length $r_0$ and calculated
the monopole+antimonopole number density $(N/V)$ as a function
of the annihilation distance $r_A$.  The results are shown
in Figs. 8 and 9 for smoothing lengths of 6 and 8 lattice spacings,
respectively.
The monopole density $(N/V)r_0^3$ decreases as a function
of $r_A$, and it must obviously go to zero as $r_A$ reaches
the size of the simulation volume.  However, it appears
that as $r_A$ approaches $r_0$, a plateau value for $(N/V)r_0^3$
is reached.
Moreover, the plateau value of $(N/V)r_0^3$ lies at the continuum
prediction of equation (\ref{numtheory}).  The statistics are rather
poor, but our results do suggest that one interpetation
of the spurious monopoles in the lattice simulations
is that such monopoles are produced as closely-spaced
monopole-antimonopole pairs.

\begin{figure}
\hspace*{1.3in}
\psfig{file=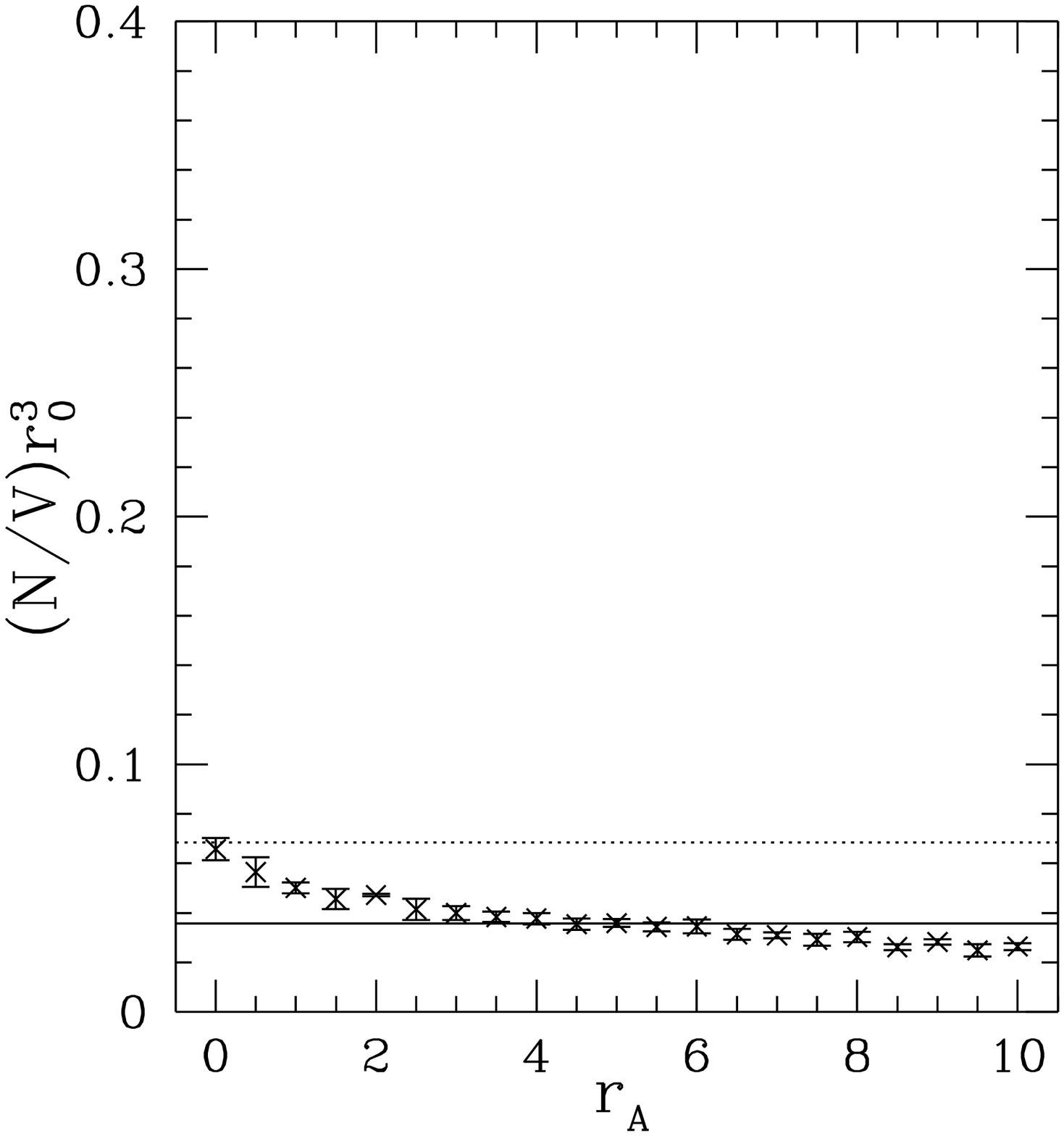,height=8.cm,width= 8.cm}
\vspace*{0.4in}
\caption{The total number of monopoles+antimonopoles
per unit volume, $(N/V)$,
multiplied by the cube of the smoothing length $r_0$,
as a function of the ``annihilation distance" $r_A$, where
monopole-antimonopole pairs separated by a distance less
than $r_A$ have been removed from the simulation,
and $r_0$ is fixed at 6 lattice spacings.
The solid line gives the analytic prediction for
$(N/V)r_0^3$ in the continuum limit.
The dotted line is the lattice-based analytic prediction.}
\end{figure}

\begin{figure}
\hspace*{1.3in}
\psfig{file=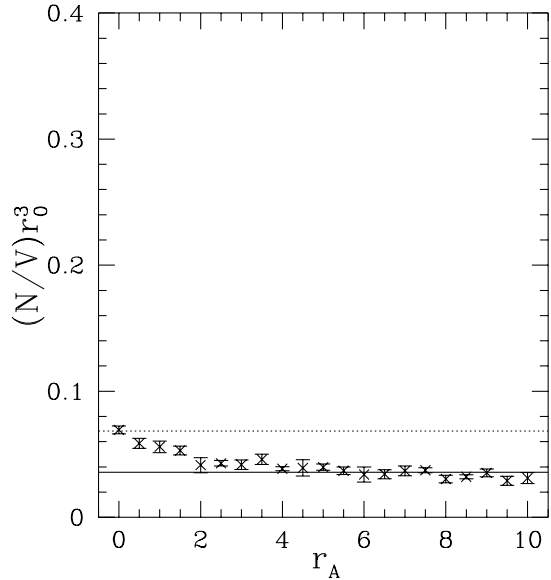,height=8.cm,width= 8.cm}
\vspace*{0.4in}
\caption{As in Fig. 8, with $r_0 = 8$ lattice spacings.}
\end{figure}

\section{Discussion}

Our results indicate that lattice simulations for
the formation of topological defects must be used
with some caution.   Even when the Gaussian fields
used in these simulations are smoothed to reduce
lattice effects on small scales, a residual lattice
dependence remains.  Another way to express this
result is to note that smoothing eliminates the
small-scale inhomogeneity introduced by using
a lattice, but it cannot eliminate the anisotropy.
Strings, for example, simulated on a cubic lattice
are required to lie only in the $x$, $y$, or $z$ directions,
while a genuine continuum simulation would allow an arbitrary
direction for the string segments.

For the case of domain walls, a simple multiplicative
factor which depends on the direction of the domain wall
can be used to eliminate this problem; all domain wall
simulations have used such a factor \cite{Press} - \cite{Larsson}.
For strings, the correct multiplicative
factor does not amount to simply averaging over the
directions for a straight string:  in the lattice
simulations there is residual small-scale
structure along the strings and probably an accompanying
distribution of tiny loops (of order the lattice spacing).
These effects change the multiplicative factor from
1.5 (its value when only anisotropy is taken into account)
to 1.91, but they should not affect the length distribution of
loops (except for the smallest loops).  In practice,
nearly all previous studies of cosmic
strings have used a simple lattice
simulation of the kind described here \cite{VV} - \cite{paper1},
as have some monopole simulations \cite{Copeland} (although
see Ref. \cite{Bennett} for a continuum field treatment of
monopoles).

For cosmic strings and monopoles,
the problem can be solved by linearly interpolating the
values of the field on the lattice.  This procedure
yields the correct continuum value for the monopole and
string densities; for the case of monopoles,
a similar result can be achieved
in a simple-minded lattice simulation
by eliminating
close pairs of monopoles.
For the case of cosmic strings, this interpolation scheme
produces a truly lattice-free estimate of the total string length
in the form of infinite strings:  $f_\infty = 0.716 \pm 0.015$.
This is nearly identical to the value obtained from a
simple lattice simulation (with Gaussian smoothing) where the
strings were constrained to lie on the edges of the lattice
\cite{paper1}.  Apparently $f_\infty$ is much less sensitive
to lattice effects than is the total string density.  This is
not surprising, since the effect of the cubic lattice
is to multiply the string density by a geometric factor;
if this factor is the same for both closed loops and infinite
strings, then $f_\infty$ is unaffected.  However, even in these
lattice-free simulations, we expect the value of $f_\infty$
to depend on the window function used
for smoothing (as in reference \cite{paper1}).

This linear interpolation procedure can also be extended
to domain walls, providing an alternative to
the weighting factor used in references \cite{Press} - \cite{Larsson}.
In practice, these two methods should give nearly identical
results for any domain wall quantities of interest.

\acknowledgements

We thank U.-L. Pen for numerous helpful suggestions, including
the linear interpolation scheme for the lattice-free string simulations.
We are grateful to
S. Larsson for helpful comments on the manuscript.
R.J.S. is supported by the Department of Energy
(DE-FG02-91ER40690) and by NASA
(NAG 5-3111) at Ohio State, and
by the Department of Energy and NASA (NAG 5-2788) at Fermilab.
A.V. is supported by the National
Science Foundation.

%
%

\end{document}